\documentclass[referee]{aa} 
\usepackage{graphicx}

\begin{document}

\title{Beryllium in turnoff stars of NGC6397: early Galaxy spallation, 
cosmochronology and
cluster formation 
\thanks{Based on observations collected at the ESO VLT, Paranal Observatory,
Chile}}

\subtitle{}
\author{L. Pasquini\inst{1}, P. Bonifacio\inst{2}, S. Randich\inst{3}, 
D. Galli\inst{3}, R. G. Gratton\inst{4}}

\offprints{lpasquin@eso.org}

\institute{European Southern Observatory, Garching bei M\"unchen, Germany
\and
INAF--Osservatorio di Trieste, Trieste, Italy
\and
INAF--Osservatorio di Arcetri, Firenze, Italy 
\and
INAF--Osservatorio di Padova, Padova, Italy }

\abstract{We present the first detection of beryllium in two turnoff
stars of the old, metal-poor globular cluster NGC~6397.  The beryllium lines
are clearly detected and we determine a mean beryllium abundance of
$\log$(Be/H)$=-12.35 \pm 0.2$.
The beryllium abundance is very similar to that of field stars of similar Fe
content.  We interpret the beryllium abundance observed as the result of
primary spallation of cosmic rays acting on a Galactic scale, showing
that beryllium can be used as a powerful cosmochronometer for the first
stellar generations. With this method, we estimate that the cluster
formed 0.2--0.3~Gyr after the onset of star formation in the Galaxy, in
excellent agreement with the age derived from main sequence fitting.
From the same spectra we also find low O (noticeably different for 
the two stars) 
and high N abundances, suggesting that the original gas was enriched in CNO
processed material.  Our beryllium results, together with the N, O, and Li
abundances, provide insights on the formation of this globular cluster,
showing that any CNO processing of the gas must have occurred in the
protocluster cloud before the formation of the stars we observe now. We
encounter, however, difficulties in giving a fully consistent picture
of the cluster formation, able to explain the complex overall abundance
pattern.
}
\authorrunning{L. Pasquini et al.}
\titlerunning{Beryllium  abundances in NGC~6397}
\maketitle


\section{Introduction}

Beryllium has a unique origin: it is produced in the interstellar
medium (ISM) by Galactic cosmic rays (GCRs) through the spallation of
carbon, oxygen and nitrogen nuclei (Reeves et al.~1970); unlike lithium
it is not produced during the Big Bang, unlike most other elements is not
produced in stars.  It is well established that $^6$Li, beryllium and B are
produced in the Galactic halo by two spallative processes:  ({\em i}\/)
collisions of accelerated C,N,O nuclei in GCRs with ISM protons and
$\alpha$-particles (a primary process leading to a linear dependence of
Be on metallicity); ({\em ii}\/) collisions of energetic protons and
$\alpha$-particles in GCRs with ISM C,N,O nuclei (a secondary process
leading to a quadratic dependence of beryllium on metallicity)
\footnote{In addition, $^{6,7}$Li can be produced also by fusion ($\alpha +
\alpha$) reactions in the ISM, a process important for the production of Li 
isotopes in a metal--poor gas}.
The primary
process ({\em i}\/) affects the Galaxy as a whole and is predicted to
dominate the production of beryllium in the metal-poor ISM of the early Galaxy
(Suzuki, Yoshii \& Kajino~1999; Suzuki \& Yoshii~2001). 
The spallative origin of beryllium implies a power-law
relation between the abundance of beryllium and the metal abundance of stars,
a trend confirmed observationally down to [Fe/H] $\simeq -3$ in
metal-poor field halo stars (Boesgaard et al.~1999, hereafter B99). If, however, one
is interested in the earliest phases of Galactic evolution, metallicity
is no longer a reliable age indicator.  As a consequence of the
dispersed character of star formation in the Galactic halo and the lack
of efficient mixing in the gas, the chemical composition of the ISM in
the first stages of Galactic evolution was strongly affected by local
enrichment produced by individual Type II supernovae.  Observations
of very metal metal poor stars (Ryan, Norris, \& Beers~1996; McWilliam~1997) 
show in fact a large, probably intrinsic,
spread of elemental abundances, interpreted as an
indication of inhomogeneous chemical evolution of the early Galaxy
(e.g. Travaglio, Galli, \& Burkert 2001, Suzuki \& Yoshii~2001).  Even if the 
most recent, high quality data show that part of this spread is due to 
the limited quality of the previous data (Cayrel et al. 2004), 
a one-to-one relation of metallicity with age in the
first stellar generations is not expected.  Conversely, beryllium, B and the
isotope $^6$Li, being produced by energetic particles generated and
transported {\it globally} on a Galactic scale, are expected to show a
much smaller spread of abundances than the products of Type II
supernovae (typically oxygen), making them ideal ``cosmic clocks'' for
dating the first stages of halo evolution (Suzuki et al.~2001, 
Beers et al.~2000).  Globular clusters  represent ideal
test cases for this new clock because they are an extremely old stellar
population for which independent, reliable age determinations are
possible (Salaris \& Weiss 2002; Gratton et al. 2003, hereafter G03).


To test beryllium as a cosmic clock it is necessary to measure beryllium in stars
which can be dated independently. Stars in old globular
clusters such as NGC 6397 are very good candidates, because their ages can
be determined in a reliable way and they have been shown to be formed
within $\sim 1$ Gyr after the Big Bang (G03).  However,
the search for beryllium in metal-poor stars has been limited so far to
relatively bright field stars ($ V\le 12.5$), because  the only
available beryllium lines are the Be II resonance doublet at 313.1 nm.  This
wavelength is very close to the atmospheric cut-off at 300 nm and the
terrestrial atmosphere heavily absorbs the incoming radiation, making
observations very challenging. In addition, the equivalent width of the
Be lines is very small and the spectral region crowded, so that high
spectral resolution is necessary.

The high UV efficiency of UVES at the VLT telescope Kueyen has opened a
new possibility, allowing for the first time the detection of beryllium in two
turn-off (TO) stars in NGC 6397. It is crucial to reach the cluster TO
because these stars have not depleted beryllium in their atmospheres.
Conversely, brighter subgiants in the same cluster show clear evidence
of Li dilution (Castilho et al.~2000, hereafter C00). Since this more fragile
element is at its original level in TO stars, then, a fortiori, 
beryllium is not expected to be depleted in their atmospheres.


\section{Basic properties of NGC6397}

NGC6397 is one of the closest and best studied globular clusters.  In
particular it has been the subject of several recent high resolution
spectroscopic studies which have shown a very good agreement on the
cluster [Fe/H] abundance and on its homogeneity along the color-magnitude
diagram. The cluster Fe abundance determined spectroscopically by
several groups (see C00 and references therein), is in
the range $-2.2 <$ [Fe/H] $<-1.8$. The most recent determinations agree
on the value [Fe/H] $\simeq -2.0$ (C00; Th\'evenin et
al. 2001, hereafter T01; Gratton et al. 2001, hereafter G01), although
these studies adopt different techniques: LTE analysis in C00
and G01, non-LTE computations in T01.  A detailed chemical
analysis of NGC6397 has been carried out by T01, who found abundances
of $\alpha$- and Fe-peak elements in agreement with those of field
stars of similar metallicity.

The Li abundance in TO stars of NGC6397 ($\log (N({\rm Li})=2.36 \pm
0.05$, Molaro \& Pasquini~1994; C00; Bonifacio et al. 2002, hereafter B02)
is consistent with the Spite's Li plateau (Spite \& Spite~1982).
The Li abundance decreases with the star advancing on the red-giant
branch (RGB) because of dilution, and only upper limits can be obtained
for stars brighter than the RGB bump (C00). The average cluster O abundance 
appears to be lower than that of field
stars of similar metallicity ([O/Fe] $\simeq 0.2\pm 0.1$ according to C00 and G01),
but variations of O are claimed among the
subgiants, possibly anticorrelated with Na (Carretta et al. 2004).
Thus, although NGC6397 is more homogeneous than most globular clusters,
it nevertheless shows signs of contamination by gas processed through
CNO cycling.

\section{Sample selection and observations}

We selected from the T01 sample the two 
brightest stars to allow the challenging beryllium observations.
The selected stars, together with their atmospheric parameters
(B02) and chemical abundances (T01), 
are listed in Table 1. 

\begin{table*}
\caption{NGC6397 stars, their atmospheric parameters and
abundances. The atmospheric parameters are from B02, 
while the abundances are from this work and T01}


\begin{tabular}{lllllll}
\hline
Star  & T$_{\rm eff}$ & $\log g$   & [Fe/H]     &  [N/H] & [O/H]   & $\log({\rm Be/H})$  \\ 
          &           &           &          &         &         &        \\
A0228     & 6274      &  4.1     & $-2.05^a$    & $-0.74$  & $-2.24$ & $-12.27$   \\
A2111     & 6207      &  4.1     & $-2.01^a$    & $-0.74$  & $-1.64$ & $-12.43$   \\
HD218502  & 6296$^b$  &  4.13    & $-1.85^b$    & $-1.95$: &         & $-12.36 $   \\
\hline
\end{tabular}

({\em a}\/) from T01

({\em b}\/) from Alonso et al. 1999
\end{table*}

The observations were carried out in service mode at the VLT
observatory in several runs during summer 2003 with the UVES
spectrograph (Dekker et al.~2000). We used a $1^{\prime\prime}$ slit
providing a resolving power of 45,000. The blue CCD was binned along
the spatial direction to minimize the CCD read out noise.  We used the
dichroic to obtain, in addition to the UV data, spectra with the red
860 nm setup, containing the O~{\sc i}~777 nm triplet which we used to
derive the O abundances.  A total of 8 exposures of 90 minutes each were
obtained per star.  The observations were reduced with the UVES
pipeline (Ballester et al.~2000).  The reduced spectra were averaged
leading to a S/N ratio in the beryllium region of 8 and 15 per pixel for star
A2111 and A228 respectively. The spectra in the beryllium region, together
with our synthetic best fits (cfr. next section) are shown in Figure
1.  In addition to the cluster stars, we obtained high S/N ($\sim 130$)
spectra of the bright reference star HD 218502, whose atmosperic
parameters are very close to the ones of the cluster stars (cfr. Table 1).

\begin{table}
\caption{Line list used for the spectrum synthesis in the region of the 
Be II lines.}
\label{lines}
{\scriptsize 
\begin{tabular}{rrrrrr}
\hline\\
Wavelength  & ion & $\log gf$ & ref. & Pred. & E           \\
(nm)        &     &           &      &       & (cm$^{-1}$) \\

\hline\\

   313.0027  & 107.00 &  -2.799 &  K     &      & 9182.250\\
   313.0056  &  40.00 &  -0.700 &  EM    &      & 4186.110\\
   313.0167  &  22.00 &  -0.468 &  K     &      &47913.551\\
   313.0190  & 107.00 &  -3.085 &  K     &     *& 9337.492\\
   313.0191  & 107.00 &  -2.657 &  K     &     *&10779.396\\
   313.0195  &  26.00 &  -3.603 &  K94   &      &28819.951\\
   313.0202  &  25.01 &  -3.714 &  K88   &      &38720.020\\
   313.0212  &  26.00 &  -2.840 &  K94   &      &58812.012\\
   313.0254  & 106.00 &  -1.180 &  K     &      & 4205.830\\
   313.0257  &  23.01 &  -0.290 &  NBS   &      & 2808.720\\
   313.0281  & 108.00 &  -1.931 &  GG    &      &    0.000\\
   313.0290  & 106.00 &  -1.136 &  K     &      & 4206.520\\
   313.0340  &  58.01 &  -0.152 &  MC    &      & 4266.397\\
   313.0353  &  27.01 &  -3.333 &  K88   &      &24074.600\\
   313.0370  & 106.00 &  -1.555 &  K     &      &  267.550\\
   313.0377  &  22.00 &  -1.594 &  K     &      &11531.759\\
   313.0419  & 106.00 &  -2.464 &  K     &      &14274.020\\
   313.0420  &   4.01 &  -0.168 &  BIE   &      &    0.000\\
   313.0439  &  25.00 &  -2.518 &  K88   &      &30419.609\\
   313.0476  & 106.00 &  -2.450 &  K     &      &14274.970\\
   313.0519  & 107.00 &  -2.701 &  K     &     *&11114.699\\
   313.0549  &  25.01 &  -1.146 &  K88   &      &52373.180\\
   313.0550  & 107.00 &  -3.586 &  K     &      & 6932.594\\
   313.0562  &  26.01 &  -5.213 &  K88   &      &30388.543\\
   313.0569  &  24.01 &  -2.457 &  K88   &      &42986.621\\
   313.0570  & 108.00 &  -1.559 &  GG    &      &    0.000\\
   313.0592  &  24.00 &  -1.968 &  K88   &      &28682.207\\
   313.0636  & 106.00 &  -1.180 &  K     &     *& 4188.995\\
   313.0637  &  25.00 &  -1.008 &  K88   &      &34423.270\\
   313.0648  & 106.00 &  -1.447 &  K     &      &  270.590\\
   313.0674  & 106.00 &  -1.136 &  K     &     *& 4189.812\\
   313.0726  & 107.00 &  -2.851 &  K     &     *& 9508.673\\
   313.0780  &  41.01 &  +0.410 &  HLB   &      & 3542.500\\
   313.0782  & 106.00 &  -1.555 &  K     &     *&  265.368\\
   313.0809  &  22.01 &  -1.140 &  MFW   &      &   94.100\\
   313.0813  &  64.01 &  -0.083 &  MC    &      & 9328.864\\
   313.0823  & 107.00 &  -2.670 &  K     &     *&11115.135\\
   313.0871  &  58.01 &  +0.481 &  CC    &      & 8789.380\\
   313.0906  & 106.00 &  -0.503 &  K     &     *&18848.234\\
   313.0910  & 107.00 &  -3.018 &  K     &     *& 9338.252\\
   313.0928  & 106.00 &  -3.136 &  K     &      &   17.770\\
   313.0932  & 108.00 &  -3.358 &  GG    &      &    0.000\\
   313.1015  &  25.01 &  -1.217 &  K88   &      &49291.309\\
   313.1037  &  25.00 &  -1.725 &  K88   &      &30425.711\\
   313.1058  & 106.00 &  -1.447 &  K     &     *&  268.362\\
   313.1065  &   4.01 &  -0.468 &  BIE   &      &    0.000\\
   313.1070  &  90.01 &  -1.559 &  MC    &      &    0.000\\
   313.1074  & 106.00 &  -0.514 &  K     &     *&18849.781\\
   313.1109  &  40.00 &  -0.400 &  CB    &      & 4196.850\\
   313.1110  &  26.00 &  -5.171 &  K94   &      &24574.652\\
   313.1115  & 107.00 &  -2.627 &  K     &     *&11113.809\\
   313.1116  &  76.00 &  +0.050 &  CB'   &      &14848.050\\
   313.1198  & 107.00 &  -3.552 &  K     &      & 6933.030\\
   313.1207  &  24.00 &  -0.659 &  K88   &      &25106.299\\
   313.1243  &  26.00 &  -4.193 &  K94   &      &17550.180\\
   313.1255  &  69.01 &  +0.280 &  AS    &      &    0.000\\
   313.1269  & 107.00 &  -3.191 &  K     &     *& 9194.144\\
   313.1326  &  27.01 &  -4.088 &  K88   &      &17771.711\\
   313.1339  &  21.01 &  -2.430 &  K88   &      &59528.422\\
   313.1395  &  26.01 &  -3.656 &  K88   &      &30764.484\\
\\
\hline\\
\end{tabular}
}
\end{table}

\section{Data analysis: Stellar parameters and determination of abundances}

Table~1 lists the adopted atmospheric stellar parameter (B02)
and the abundance values determined in this work. 


A full synthesis of the region around the beryllium doublet has been performed
using the line list given in Table~\ref{lines}.  In this table the ions
are identified by their code, according to Kurucz's convention (Kurucz
1993).  Almost all  lines have been extracted by a version of  the
Kurucz data base updated by F. Castelli (private communication).  The
only exception are the OH lines, for which gf values have been computed
from lifetimes of Goldman \& Gillis (1981).  We used our Linux
version (Sbordone et al. 2004) of the SYNTHE code (Kurucz~1993).  For
each star the best abundance was found by a $\chi^2$ fit to the whole
feature. The best fits, shown in Fig.~1, correspond to beryllium abundances of
log(Be/H)$=-12.27$, and (Be/H)$=-12.43$ for A228 and A2111,
respectively.  For HD218502, we derive an abundance of
$\log$(Be/H)$=-12.36\pm 0.1$, in agreement with the previous
determination by Molaro et al.~(1997), $\log$(Be/H)$=-12.56 \pm 0.22$.

We estimate the errors with Monte Carlo simulations assuming a Poisson
noise.  The dispersion around the mean in 1000 Monte Carlo samples was
0.09 dex for A2111 and 0.11 dex for A228. These should be viewed as
lower limits on the errors, since at these low S/N ratios other sources
of non-Poisson noise (e.g.  shot-noise) could be important. To these errors
associated to the noise in the data one must add the errors which
derive from the uncertainty in the atmospheric parameters.  Since we
are using lines of singly ionized beryllium, surface gravity is the parameter
which most affects the beryllium abundance computation.  The values of $\log
g$ have been estimated from the position of the stars in the
color-magnitude diagram and theoretical isochrones.  
Since the uncertainty in the derivation of gravity in the case of these 
clusters stars is dominated by the unceratinty in their mass, we can safely assume an
error of 0.15 dex in $\log g$ that translates into an
uncertainty in the beryllium abundance of 0.08 dex.  The error arising from a
change of 100 K  in the effective temperature is also
0.08 dex. Summing these two errors under quadrature one obtains 0.11
dex.  If we add in quadrature this value to the errors due to noise
(0.09~dex and 0.11~dex, respectively), we obtain an estimate of the
total error on the beryllium abundance of 0.14 dex for A2111 and 0.15 dex for
A228.  Therefore we conclude that the beryllium abundance of the two stars is
the same within the errors. We can then estimate the average beryllium
abundance of NGC6397 as $\log$(Be/H)$=-12.35\pm (0.10)_{\rm stat} \pm
(0.11)_{\rm sys}$.  The systematic error is  estimated as due to an
uncertainty of the zero point of the temperature scale by 100 K and an
uncertainty of the zero point of the $\log g$ scale by 0.15 dex.  
Note that this error does not include possible systematic errors
due to shortcomings and inadequacies of our modelling (model
atmospheres, atomic data, etc.). 

\begin{figure}[ht]
\begin{center}
\resizebox{14cm}{!}{\includegraphics{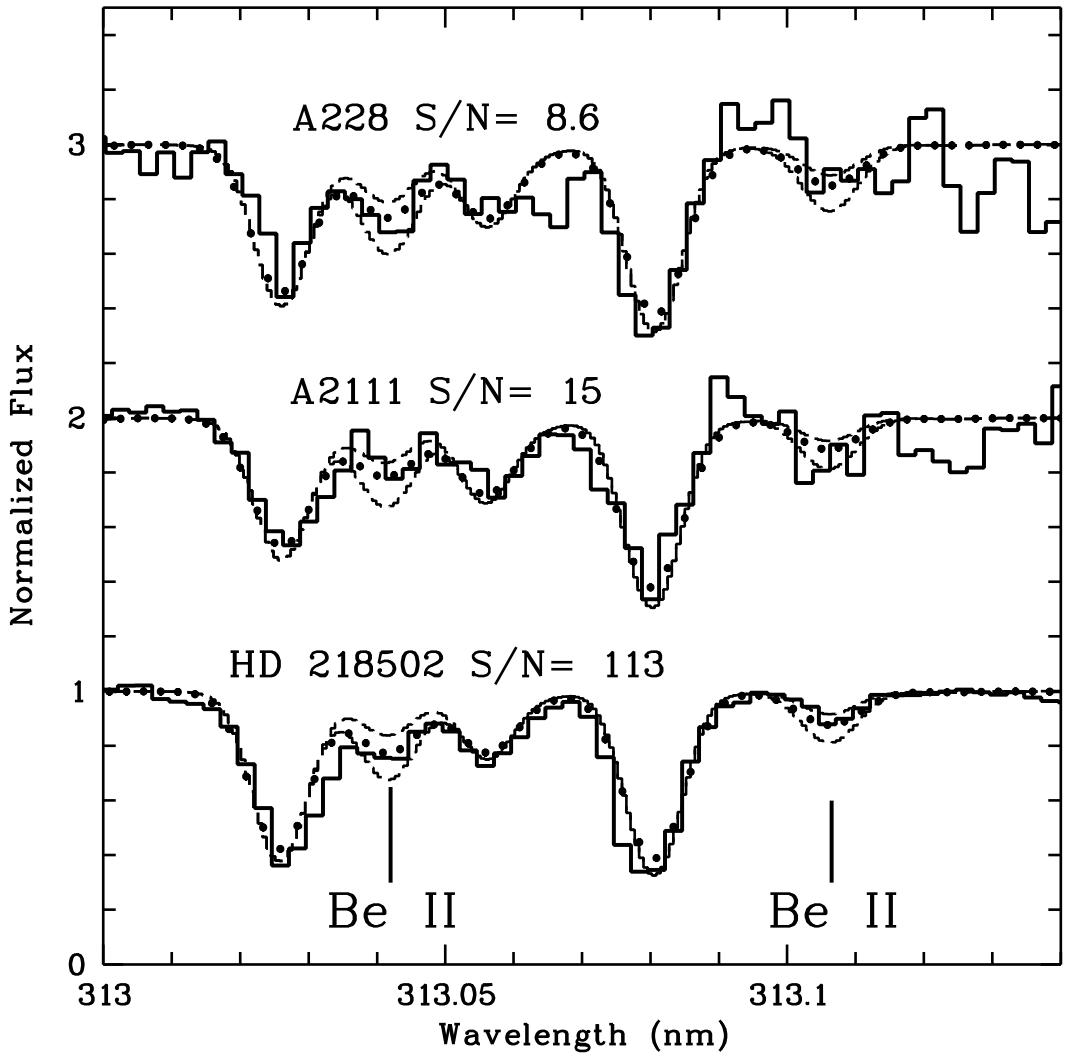}}
\end{center}
\caption{UVES spectra of the two turn-off stars A228 and A2111 of the
globular cluster NGC~6397. The dotted lines correspond to the best-fit
Be abundances $\log$(Be/H)$=-12.27$ for A228 and $\log$(Be/H)$=-12.43$
for A2111. The  dashed lines correspond to synthetic spectra
with abundances of $\pm 0.2$ dex from the best fitting abundance. 
For comparison, we also show the spectrum of the bright halo
star HD 218502, whose atmospheric parameters are close to those of the
NGC~6397 TO stars.  The best fit synthetic spectrum for this star
corresponds to $\log$(Be/H)$=-12.36$.
}
\end{figure}

The oxygen abundance has been computed from  the infrared triplet
lines, using the same method as G01. In Figure 2 we
show the spectra of the two NGC~6397 stars and of HD218502; the triplet
lines are very weak, and in particular for A228 they are at the limit
of detectability. Considering that the two cluster stars have similar
stellar parameters, Figure 2 suggests that the two stars have different
O abundances.  In fact we derive [O/H]$=-1.64\pm 0.05$  and
[O/H]$=-2.24\pm 0.15$ for A2111 and A228, respectively. Previous
studies of O in stars of NGC~6397 have obtained low O abundances
([O/Fe]$\simeq 0.2$ according to C00 and G01) with
respect to stars of similar metallicity. Carretta et al.~(2004) have
shown that a spread in O abundance is present among the subgiants of
this cluster. For the first time evidence of O dispersion is also found
in main sequence stars of NGC~6397.

\begin{figure}[ht]
\begin{center}
\resizebox{14cm}{!}{\includegraphics{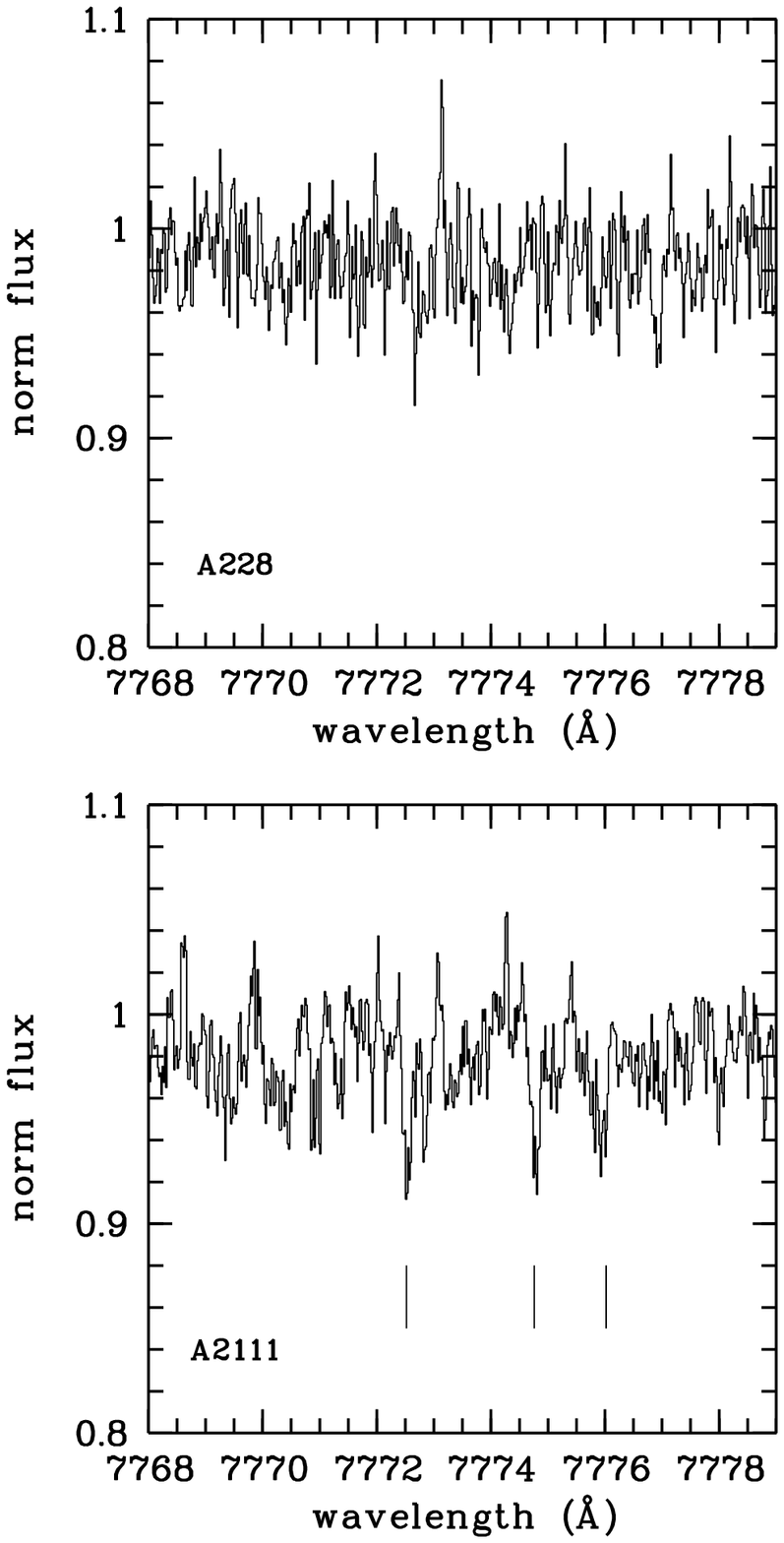}}
\end{center}
\caption{UVES spectra of stars A2111 and A228 in the oxygen triplet region. The two
stars have identical atmospheric parameters, but the O abundance
differs almost by 0.6 dex.} 
\end{figure}

We have also been able to measure the N abundance in these two cluster
stars, confirming that the TO stars are N-rich, like the subgiant stars
studied by Carretta et al.~(2004).  Nitrogen is a difficult element to
measure, especially for dwarf stars, since the NI lines become extremely
weak at moderately low metallicity.  On the other hand, CN bands, which
can be used down to very low metallicity in giant and sub-giant stars
are not measurable in dwarfs.  At low metallicities the only feature
which has been used to measure N is the NH band at 336 nm, which is
very near to the atmospheric cut-off. Among field stars the [N/Fe] ratio is
approximately constant with metallicity in the range $-3.0 \le
$[Fe/H]$\le 0.4$, as first shown by Carbon et al. (1987) and recently
confirmed by Israelian et al.~(2004) and Ecuvillon et al.~(2004) on the
basis of higher quality data.  We notice however the existence in these
samples of a non-negligible population of N-rich stars, i.e. stars
which have a N abundance higher than most stars of similar
metallicity.  The presence of a few N-rich stars has been known for a
long time (Bessell \& Norris 1982; Laird 1985) but their census is
steadily increasing.  


We estimate the nitrogen abundance by spectral synthesis of two NH UV
bands, around 336 nm and around 340.5 nm.  Given the complexity of the
bands it is not possible to carry out a formal fitting procedure of the
whole spectral region. We have therefore inferred a ``best abundance''
by fitting the most prominent features.  Despite several systematic and
random uncertainties, we are able to confirm the overabundance of N in
both NGC6397 TO stars with respect to N-normal field stars of similar
metallicity.  In Fig. 3 we compare the NH bands of the two NGC~6397 stars  and of
HD218502, in order to emphasize the N-rich nature of the cluster stars.

\begin{figure}[]
\begin{center}
\resizebox{14cm}{!}{\includegraphics{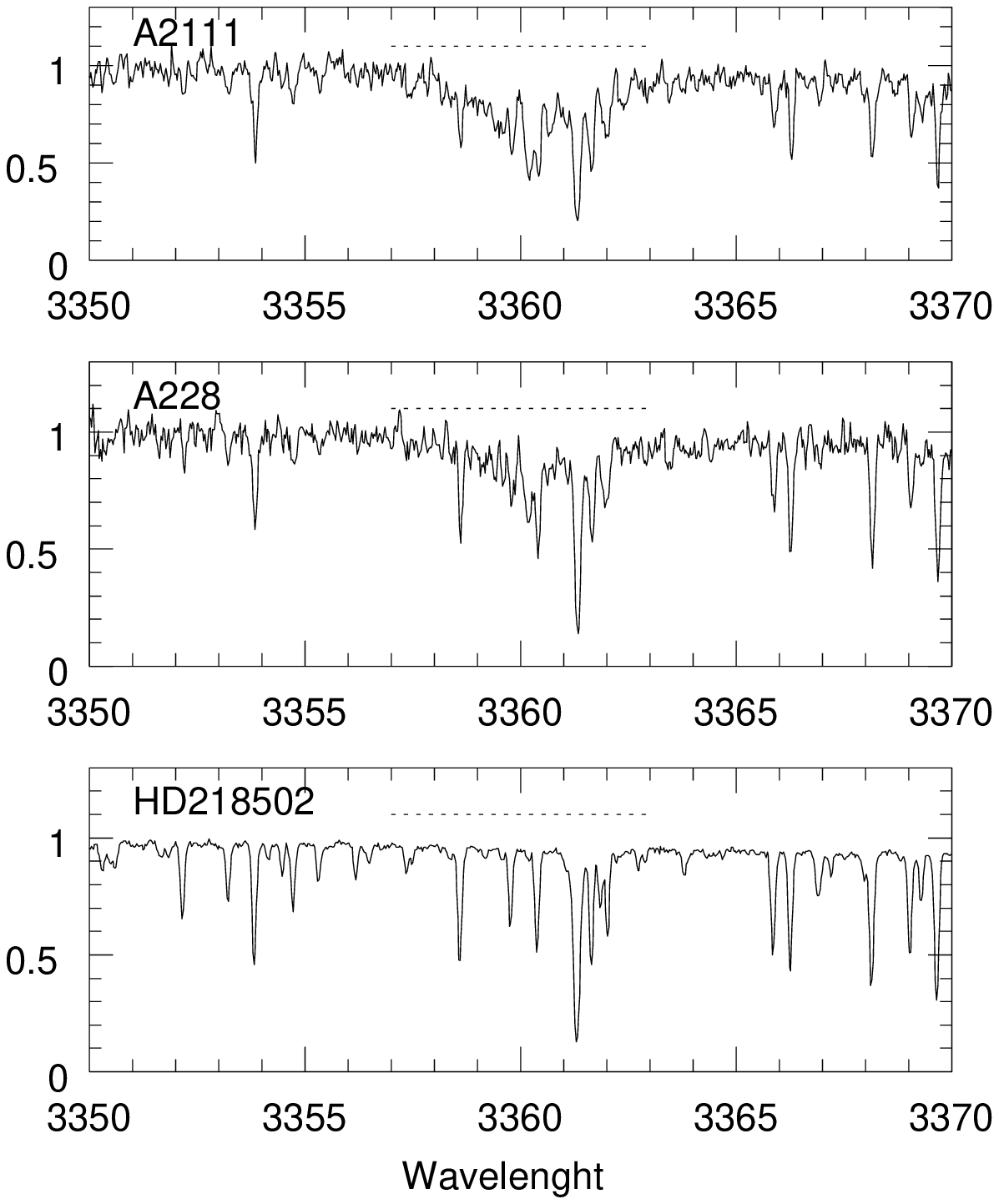}}
\end{center}
\caption{UVES spectra of stars A2111 and A228 around  the NH band at 
336~nm. For comparison, we also show the same spectral region for our
reference star HD218502, whose atmospheric parameters ($T_{\rm
eff}=6296$~K, $\log g=4.13$, [Fe/H]$=-1.85$) are close to those of the
NGC6397 TO stars.  While the band (whose core region is indicated by the dashed lines)
is virtually absent in the reference
star, it is well pronounced in the two cluster stars,
indicating a strong N overabundance.} 
\end{figure}


\section{Implications on Galactic chemical evolution and cluster formation}

In this section we discuss the theoretical implications of our beryllium
determination in NCG~6397. We first show how the beryllium abundance in halo
stars can be used as a powerful cosmochronometer, due to the fact
that spallation in the early Galaxy was dominated by primary
fragmentation on a Galactic scale. We then discuss the complex
abundance pattern observed in NGC~6397 and its implications 
on cluster formation scenarios.


\subsection{Beryllium vs. Oxygen in cluster and halo stars}

As mentioned in the Introduction, cosmic-ray spallation in the early
Galaxy was dominated by primary reactions, where fast, heavy (mostly
CNO) nuclei in the cosmic rays are broken by encounters with ISM
protons and $\alpha$ particles. This process predicts a linear
dependence of beryllium with metallicity in the whole Galaxy, as well as an
approximately linear increase of beryllium with time.  One of the most
important characteristics is that, being dominated by a global
production mechanism, the beryllium abundance is expected to be largely
independent of the local chemical inhomogeneities which may have been
present in the early halo (Beers et al.~2000, Suzuki et al. 1999).
On the other hand,  the cosmic ray flux is driven by the global star formation 
rate and to a certain extent by the confining mechanism, which were likely 
related to the large scale magnetic fields, and our knowledge of all these 
early quantities is very limited. 

The contribution of secondary spallation processes (collisions of
accelerated protons and $\alpha$ particles with heavy nuclei in the
ISM) is not dominant at these early epochs, and therefore the observed
underabundance of O in the cluster stars should not have dramatically
affected the beryllium abundance. In any case, since the beryllium production factors
from spallation of C,N,O nuclei are very similar (cfr. e.g.  Valle et
al.), C or N overabundances may compensate an O deficit and result in a
comparable amount of beryllium as from a ``standard'' mixture.

The beryllium abundance of the NGC~6397 stars is fully consistent with the
well-known Be vs. Fe trend (Gilmore et al. 1992, Molaro et al. 1997, B99).  In Figure 4 we
show the Be vs. O relation for the field stars from B99, and the
NGC~6397 values determined in this work. Although the O determination
is carried out in different ways in our and Boesgaard's study, a
comparison of 13 stars in common with G03 shows an excellent agreement between
the two scales.
As Figure~4 shows, the cluster stars lie above
the mean Be vs. O trend of field halo stars, as expected since they
have lower [O/Fe] than field stars with the same Fe content.  In
particular, Figure~4 shows that whereas A2111 is marginally consistent
with the general Be vs. O trend in field stars, A228 is strongly
oxygen-poor given its beryllium content.  This discrepancy can be understood
if primary spallation mechanisms dominate the beryllium production in the
early Galactic phases. In that case, the amount of beryllium produced depends
more on the cosmic-ray composition than on the local composition of the
protocluster gas.


\begin{figure}[ht]
\begin{center}
\resizebox{14cm}{!}{\includegraphics[angle=270]{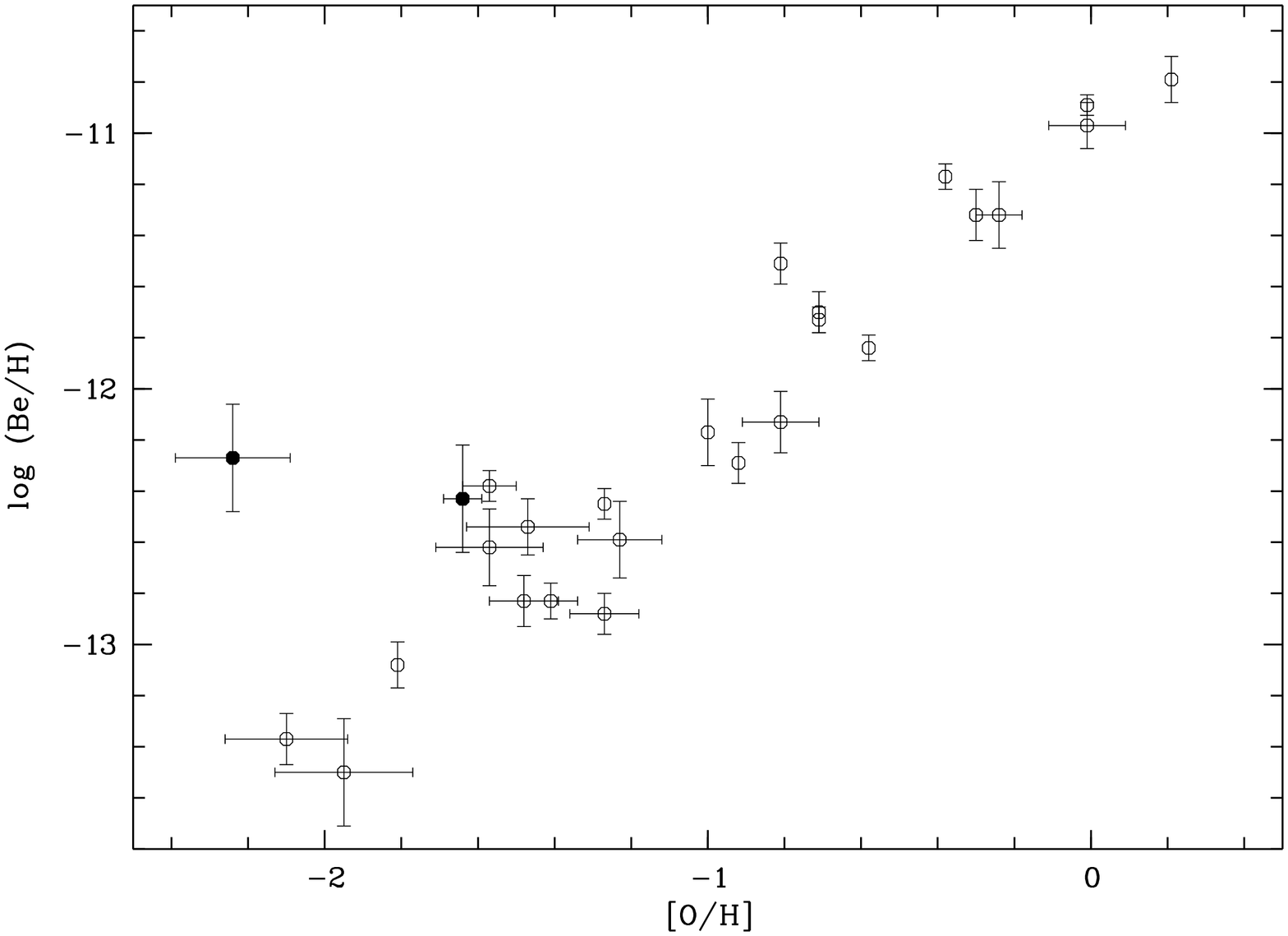}}
\end{center}
\caption{Be abundance vs. oxygen 
for the NGC6397 TO stars ({\it filled circles}) and the star from B99,
({\it  open circles}).}
\end{figure}


\subsection{Be as cosmochronometer and the age of NGC~6397}

NGC~6397 is an ideal target to prove the usefulness of beryllium as a cosmic
clock, as outlined in the Introduction.  This cluster has been
independently dated via main-sequence fitting and theoretical
isochrones (G03) and its TO stars are bright enough to allow detection
of beryllium in their atmospheres. Using standard isochrones, G03 find that
the age of the cluster is $13.9 \pm 0.4$ Gyr. Taking into account a
small amount of gravitational settling in the stellar tracks, reduces
the age of the cluster by about 0.5 Gyr.  Therefore, the best age
estimate for this cluster is $13.4 \pm 0.8 \pm 0.6$~Gyr (G03).

Figure 5 shows the expected evolution of beryllium with time, resulting from a
model of chemical evolution (Valle et al.~2003) following the
enrichement of three different regions in the Galaxy, coupled by mass
flows: the halo, the thick disk and the thin disk.  The beryllium abundance of
the model is normalized to the solar meteoritic value, assuming that
the solar system formed 4.5~Gyr ago.  The other data points in Figure~5
shows the beryllium abundance in the young open cluster IC2391 (Randich et al.
2002), as indicator of the present day value, and the average value of
the two stars of NGC~6397, determined in this paper.  We have assumed a
Galactic age of 13.7~Gyr, based on WMAP data (Bennett et al. 2003).
This value represents the ``age of the Universe'' (time elapsed since
the Big Bang), whereas the actual Galactic age (time elapsed since the
onset of star formation in the Galaxy) should take into account the
time interval between the Big Bang and the epoch of reionization. The
best estimate for this interval, according to WMAP data, is $\sim
0.18$~Gyr, corresponding to $z_{\rm reion}=20$, within the errors on
the age estimate of NGC~6397.

\begin{figure}[ht]
\begin{center}
\resizebox{9cm}{!}{\includegraphics{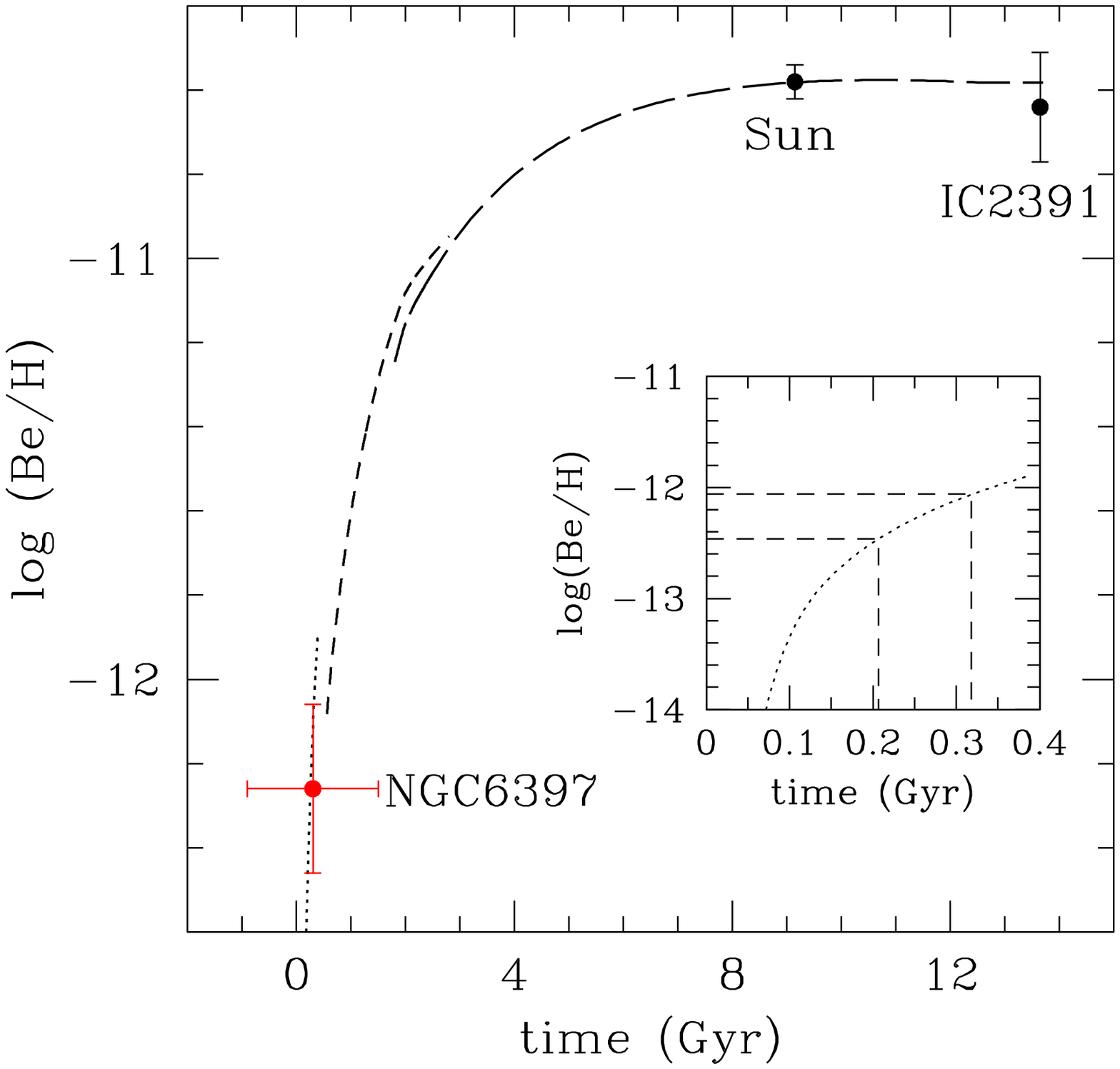}}
\end{center}
\caption{Evolution of beryllium with time in the Galaxy according to a
three-zone Galactic chemical evolution model (Valle et al. 2002). The
three curves refer to the halo, thick disk, and thin disk. The data
points show the beryllium abundance in the young open cluster IC2391 (age
$30$~Myr), the Sun (age $4.5$~Gyr), and the globular cluster NGC6397. 
The model result is normalized to the solar meteoritic
abundance. The inset illustrate the use of beryllium as a ``cosmic clock'' to
constrain the formation of NGC 6397. The horizontal  
lines corresponds to the 1 $\sigma$ counturs around the measured beryllium 
abundance.The cluster birth is constrained to the first 0.2--0.3~Gyr after
the onset of star formation in the Galactic halo. 
}
\end{figure}

The inset in Figure~5 shows the halo evolution of beryllium on a finer scale,
with the two horizontal lines limiting the 1$\sigma$ range of beryllium
abundance in NGC6397.  This plot emphasizes the possible use of beryllium as
``cosmic clock'':  the measured value of beryllium indicates that the
formation of NGC6397 occurred about 0.2-0.3 Gyr after the onset of star
formation in the Galactic halo. Considering all the uncertainties
present in the model it is safe to  conclude that the birth of the
stars composing NGC~6397 took place within the first $\sim 0.5$ Gyr of
the halo evolution, in agreement with the isochrone fitting age
determination for this cluster. We stress the fact that the ``Be age''
of the cluster is a measure of the time interval between the onset of
star formation in the Galaxy and the formation of the cluster, whereas
the ``evolutionary age'' of NGC~6397 is a measure of the time elapsed
since the formation of the cluster and the present epoch.  The agreement
of the two methods indicated a global consistency of the results of
stellar evolution, cosmic-ray nucleosynthesis and cosmology.  This
result is largely independent from the chemical model adopted, because
for the early phases of Galactic evolution the beryllium production rates
between different models agree to within better than a factor 2, 
which is comparable to the uncertainties of the beryllium
determination in NGC6397.

The ``beryllium age'' method can be extended in principle to field halo stars
for which independent age determinations are difficult (see e.g. Beers
et al.~2000).  We note that according to this approach, and under the
hypothesis that the non-zero metallicity field stars we observe today
formed several Myr after the onset of the star formation in the halo,
then their beryllium should be at  low, but  non-negligible  levels of
abundance (e.g. $\log$(Be/H)$\simeq -13.6$ for 100 Myr). 
On the other hand their metal abundances should be sensitive to local enrichment
phenomena, thus stars of the same age and beryllium abundance could show
different [Fe/H] and [O/H]. 
The beryllium-metallicity relationship may therefore present
some considerable scatter at the lowest metallicity end, as possibly
indicated by recent observations of extremely metal poor stars (B99, Primas et
al. 2000).

%

\subsection{The formation of NGC 6397} 


As we have already discussed in Sect.~4, recent observations of
NGC~6397, as well as our own results, strongly indicate an oxygen
underabundance with respect to field stars with similar Fe (C00;
G01), a nitrogen overabundance, and a star-to-star scatter in
oxygen abundance (Carretta et al.~2004). Other $\alpha$ and Fe-peak
elements (like Mg, Na, Ca, Sc, Ti, Cr, Ni, Zn, and Ba), however, do not
show significant star-to-star variations, and are consistent with those
of field halo stars of similar metallicity (G01, T01).
The low O and high N abundances suggest that the stars of NGC~6397 were
either formed from, or were partially polluted by, material bearing the
signature of asymptotic giant-branch (AGB) products (Chieffi et al.
2001; Ventura, D'Antona \& Mazzitelli~2002). In this phase, stellar material is
processed at the very high temperatures ($\sim 10^8$~K) where O is
effectively burnt to N in the CNO cycle and then returned to the ISM by
mass loss.  However, at the high temperatures characterizing AGB
nucleosynthesis, Li and beryllium are completely destroyed. It is therefore
quite challenging to explain the normal (Pop II) level of Li observed
in NGC~6397 (B02). In principle, Li could be produced during the AGB
phase and brought to the surface through the Cameron-Fowler mechanism
(Cameron \& Fowler~1971, Sackmann \& Boothroyd~1992). However, a
remarkable fine tuning between Li production and destruction is needed
to bring at the surface of these AGB stars (and into the ISM) an amount
of Li exactly equal to the primordial value (see discussion in B02 and
references therein).  Another weak point of this scenario, is that
according to our observations, at least some of the low-mass stars of
NGC~6397 should have been formed from material partially polluted by a
previous generation of AGB stars. Just as an example, if we assume that
the star A2111 is representative of an unpolluted star, the O abundance
of star A228 indicates that $\sim 3/4$ of its mass were polluted by AGB
ejecta. In addition, the high N content of star A2111 (which is supposely an example of 
'unpolluted' object) will remain unexplained. 
The presence of beryllium,  which is only
destroyed during the AGB phase, and is not produced by stellar
nucleosynthesis, makes  the possibility that a {\it
considerable fraction} of the gas of the protocluster was processed by
a previous generation of AGB stars quite unlikely. 

One could image an AGB scenario where  the gas was processed 
extremely early in the AGB stars and immediately released to the ISM, 
where it was exposed to the GCRs, but it seems difficult to 
satisfy all the previous points. Assuming a different perspective,  one could  
argue that  N rich gas is primarily produced in massive stars 
and then that N-rich, H-poor gas is released to the ISM
(see e.g. Meynet \&Maeder~2002 for possible yelds). 
While this scenario could explain the N overabundance,  it 
would require some additional  mechanism to explain the 
origin of the O underabundance.

We notice the similarity of the timescales involved in the 
different enrichment processes probably at work in these early phase of
halo evolution. While the lifetime of a 5~$M_\odot$ star (a typical AGB
progenitor) of zero metallicity is $\sim 0.1$~Gyr (Chieffi et al.
2001), our model of Galactic evolution shows that an irradiation
timescale of $\sim 0.2$~Gyr is required to bring the beryllium abundance up to
the observed values. According to our model, a similar timescale is
also necessary to raise the iron content of the gas from zero to
[Fe/H]$\simeq -2$.  Thus, the emerging picture seems to indicate the
simultaneous presence in the early Galaxy of both local (Type II
supernovae, AGB stars) and global (GCRs) enrichment processes, acting
on comparable timescales during the rapid phase of star formation in
the halo.

We are therefore left without a satisfying solution to the overall
abundance pattern observed in NGC~6397. Our results on beryllium, however,
imply that the TO stars we observe today did not process material 
in their interiors and their chemical composition is representative
of the primordial protocluster gas.

\begin{acknowledgements}
We thank A. Renzini and J. Walsh  for several comments and suggestions. 
SR and DG acknowledge financial support by the Italian Ministero dell'Istruzione, dell'Universit\`a e 
della Ricerca through a COFIN grant. Finally, we 
would like to thank the UVES team for building such a 
wonderful instrument. 
\end{acknowledgements}

\end{document}